\documentclass[doublecol]{epl2} 

\usepackage{epsfig}
\usepackage{latexsym,amsmath,amssymb,amstext}
\usepackage{float}
\usepackage{graphicx}
\usepackage[normalem]{ulem}
\usepackage{color}

\newcommand\bef{\begin{figure}}
\newcommand\eef[1]{\label{fg:#1}\end{figure}}
\newcommand\beq{\begin{equation}}
\newcommand\eeq[1]{\label{#1}\end{equation}}
\newcommand\beqa{\begin{eqnarray}}
\newcommand\eeqa[1]{\label{#1}\end{eqnarray}}
\newcommand\bet{\begin{table}}
\newcommand\eet[1]{\label{tb:#1}\end{table}}

\newcommand\fgn[1]{Figure \ref{fg:#1}}
\newcommand\eqn[1]{eq.\ (\ref{#1})}

\newcommand\ie{{\sl i.e.\/}}

\newcommand\pr{{\sl Phys.\ Rev.\/}\ }

\newcommand{\mloc}{m_{\rm loc}}

\title{Ground state mass in short lattices by controlling overconfidence and bias in Bayesian fits}
\shorttitle{Ground state mass in short lattices by Bayesian fits} 

\author{Sourendu Gupta\inst{1} \and Anirban Lahiri\inst{2}}
\shortauthor{Sourendu Gupta \& Anirban Lahiri}

\institute{                    
  \inst{1} Department of Theoretical Physics, Tata Institute of Fundamental Research, \\ Homi Bhabha Road, Mumbai 400005, India. \\
  \inst{2} Fakult\"at f\"ur Physik, Universit\"at Bielefeld, D-33615 Bielefeld, Germany.
}

\pacs{11.15.Ha}{Lattice gauge theory}
\pacs{02.50.Tt}{Inference methods}
\pacs{12.38.Gc}{Lattice QCD calculations}

\abstract{We investigate the seemingly ill-defined problem of extracting a
ground-state mass from a lattice simulation where the extent of the
lattice is not long enough to project out the ground-state properly.
We regulate the problem using a Bayesian method. We show that controlling
meta-parameters (overconfidence) can allow the data to overcome the
input priors (bias). We can write the method as a black-box technique
which allows extraction of a ground-state mass, even on a relatively
short lattice.
}

\begin{document}

\maketitle

\section{Introduction}\label{sec:intro}

The numerical extraction of physically relevant quantities from simulations
of lattice gauge theories typically involve fitting. The paradigmatic
statistical problem is to extract the masses from the measurement of a
correlator using the fitting formula
\beq
   C(t) = \sum_{i=0}^\infty A_i \cosh\left[m_i\left(\frac N2-t\right)\right],
\eeq{fitfn}
where we assume an ordering $m_0<m_1<m_2\cdots$, and the Euclidean
time $t$ runs over a finite set of integers, $0\le t< N$, and $N$ is the
lattice extent in the direction of Euclidean time.  The infinite number of
parameters is effectively truncated because all the terms with $1/m_i\le1$
can be absorbed into a single mass $m_M$ and coefficient $A_M$, so
that the sum in \eqn{fitfn} can be taken to run only over $M$ states.
We will examine the case where periodic boundary conditions are applied,
although other boundary conditions can be dealt with by a straightforward
extension of the analysis we present.

In the usual fitting method one strives to take $N\gg1/(m_1-m_0)$. In
that case there is a long interval of $t>T$ where one term of the fit
suffices. This is usually monitored by computing local masses, $\mloc$,
which are defined as the solution of the equation
\beq
    \dfrac{\cosh[\mloc(t-\frac N2)]}{\cosh[\mloc(t-1-\frac N2)]}
    = \dfrac{C(t)}{C(t-1)},
\eeq{lmass}
where the correlation functions on the right are measured inputs. For
those $t$ where $\mloc$ is constant and independent of $t$, the ground
state dominates and the value of $\mloc$ is an estimate of $m_0$. It is
clear that very close to $t=N/2$ and $t=0$, the expression for $C(t)$
in \eqn{fitfn} cannot be dominated by the $i=0$ term. As a result, the
plateau in $\mloc(t)$ that we would like to observe cannot be close to
either end. So one must search for $N\gg T\gg 0$, which usually implies
that simulations must be performed with large $N$.

Some time ago a Bayesian method was introduced \cite{bayes} which did
not search for $T$. With sufficient control over the method, one
could think of extracting the ground state mass even when the condition
$N\gg1/(m_1-m_0)$ is violated, and $\mloc$ is not constant. Since the
CPU cost for a simulation grows a little faster than linearly in $N$
(every other parameter being fixed), it would be useful to understand the
fitting process well enough to be able to reduce $N$ with confidence. In
a recent study of hadron masses extracted from QCD simulations of two
flavours of light staggered quarks, we came across one such case.

In this paper we examine the process of Bayesian fitting to decide
between various ways of treating meta-parameters. Although the process
we finally use, successfully, has been used earlier qualitatively
\cite{jaynes,bishop}, there has been no quantitative statement before. In
fact, radically different treatments of meta-parameters have been used
in lattice gauge theory \cite{other}. So we feel it is important to set
down this method as we have used it.

\section{Understanding the method}\label{sec:coins}

\subsection{A simple model}

Perhaps the simplest problem of statistical inference is the extraction of
the probability that a tossed coin will land heads up. The frequentist
answer is to observe tosses of the coin $N$ times. If the number of
times this lands heads up is $H$, then the frequentist answer is
\beq
   p = \frac HN \left[1 + {\cal O}\left(\frac1{\sqrt N}\right)\right].
\eeq{limits}
Understanding the error term is the beginning of a sophisticated analysis
\cite{feller}, which is common background knowledge today for physicists.

A very small number of experiments, $N$, may by chance yield extreme
values of $p$ such as 0 or 1. If one analyzes the error terms seriously
then one will assign large errors to these extreme values. However,
it is possible to regulate these extreme values by doing a Bayesian
analysis instead.

A Bayesian analysis of the same experiment would begin by noting that one
could improve the method of inference by bringing our prior knowledge
into the analysis \cite{jaynes}.  Since all of us have observed tosses
of coins before, we should take this prior knowledge into account. Call
$p_0$ our prior knowledge of the probability. This comes from some
previous observation which we may or may not have been scrupulous
about recording.  We can quantify the depth of our prior knowledge by
a number $N_0$. The quantity $N_0$ has no experimental significance
at all, since we have never recorded any previous measurements, and
therefore, corresponds to what we may call meta-parameters. Quantities
such as these are also sometimes called nuisance parameters, since
they are not values of parameters we are interested in extracting.
Note also, that the meta-parameters are necessary in order to combine
two different experiments. In that case $N_0$ is simply the number of
coin tosses made in the first experiment.  An effective prior count of
heads is $H_0=p_0 N_0$. Note that the choices of $p_0$ and $N_0$ are
completely arbitrary.

The experiment adds to our knowledge, so the result should give us
\beq
   p = \frac{H_0+H}{N_0+N}
\eeq{prior}
To analyze how rapidly the experiment changes our prior knowledge, 
we first note that the prior leads us to expect that the result of the
experiment should be $E_0[H]=Np_0$ heads. Suppose that the prior knowledge is
very weighty, \ie, $N_0\gg N$. Then one may write
\beq
   p = p_0 + \frac{H-E_0[H]}{N_0} \left[1 - \frac N{N_0} + \frac{N^2}{N_0^2}
         +\cdots \right],
\eeq{largebias}
so, in this case, the effect of the experiment shifts our experience
only slightly. If, on the other hand, our previous experience is slight,
so that $N_0\ll N$, then
\beq
   p = \frac HN + \left(\frac{E_0[H]-H}N\right)\,\left(\frac{N_0}N\right)
    \left[1 - \frac{N_0}N + \frac{N_0^2}{N^2} + \cdots\right].
\eeq{smallbias}
In this case, our previous experience adds mildly to the results of
the experiment.
The leading term is the frequentist result, but note that the subleading
term starts as $1/N^2$, instead of $1/N^{3/2}$. In the special case $H=0$,
which needed regulation, the leading result is not $p=0$, but $p=p_0 (N_0/N)^2$.

An instructive way to understand this is to define $p_1=H/N$ and $n=N/N_0$. Then the
Bayesian formula becomes
\beq
   p = p_1 + \frac{p_0-p_1}{1+n}.
\eeq{general}
The cross over from the region where $n$ is small and \eqn{largebias}
holds, to that where $n$ is large and \eqn{smallbias} holds becomes
clear. The region $n\simeq1$ is the region of cross over. The Bayesian
point of view is often accused of introducing a bias in the experiment in
the form of a value for $p_0$. This is clearly true, and, to the Bayesian,
a feature and not a bug. However, it is clear from the expression in
\eqn{general} that the problem of bias lies essentially in the choice
of $N_0$, \ie, in deciding the relative weights given to the prior
and the experiment. In order to pin down this notion, we may say that
bias (through prior assumptions for physical parameters) is not as
important as the degree of overconfidence (through inappropriate choice
of meta-parameters).

This discussion can be phrased in the language of fitting parameters
to the results of an experiment.  This closely parallels a discussion
in \cite{bishop}. Use the convention that heads are recorded as 1 and
tails as 0.  The experiment generates a series of 0s and 1s, ${\cal E}
= \{x_i | 1\le i\le N, x_i \in \{0,1\}\}$.  Clearly
\beq
   H = \sum_i^N x_i.
\eeq{heads}
The probability distribution from which the $x$s are drawn is
\beq
\begin{aligned}
   B(x_i|p) &=& p^{x_i}(1-p)^{1-x_i}, \qquad{\rm and}\qquad \\ 
   P({\cal E}|p) &=& \prod_{i=1}^N B(x_i|p) = p^H(1-p)^{N-H},
\end{aligned}
\eeq{bernoulli}
since the successive tosses are independent. Note that $p$ is assumed
to be given.  The expected value of $x$ is $E[x]=p$. The variance is
$V[x]=p(1-p)$.  The frequentist method of extracting the parameter $p$
form the data $\cal E$ would be to maximize $\log P({\cal E}|p)$. This
gives the result $p=H/N$.

\bef
\begin{center}
\includegraphics[scale=0.30]{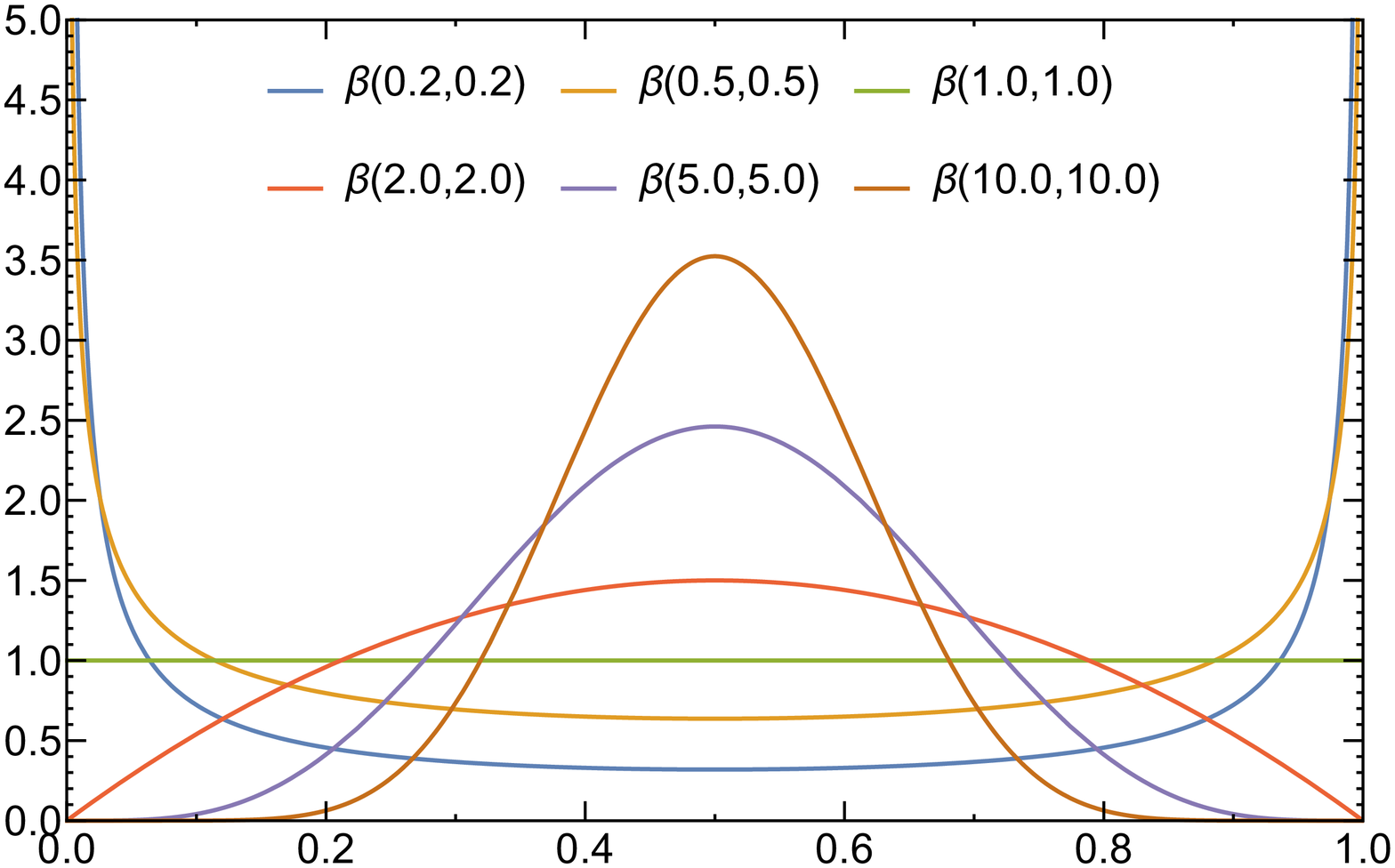}
\vskip 0.1 in
\includegraphics[scale=0.30]{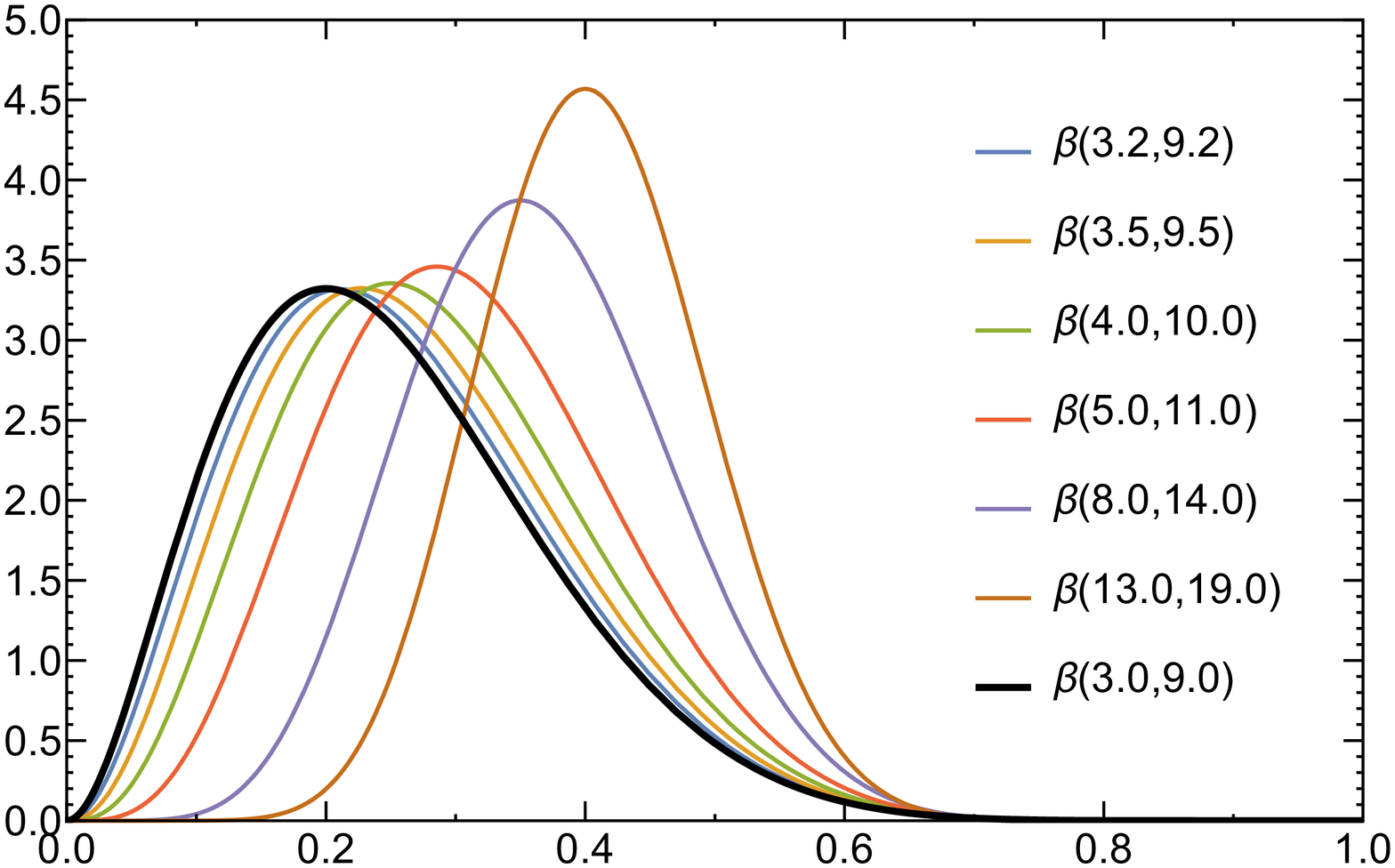}
\vskip 0.1 in
\end{center}
\caption{Several prior distributions (upper panel), all with $p_0=1/2$ and different
 $N_0$. The x-axis is $p$ and the y-axis is the probability distribution. With
 increasing $N_0$ the prior distribution is narrower. The corresponding
 posterior distributions (lower panel), with $H=3$ and $N=12$ so that $p_1=1/4$.
 For small $N_0$, the posterior distributions $P(p|{\cal E})$ are close to
 the thick black curve, which is $P({\cal E}|p)$, \ie, the likelihood function.}
\eef{experiment}

The Bayesian approach is to find the probability distribution
of $p$ given $\cal E$. By Bayes' theorem, one has
\beq
   P(p|{\cal E}) \propto P({\cal E}|p)\,{\rm Pr\/}(p),
\eeq{eeq}
where ${\rm Pr\/}(p)$ is the prior distribution of $p$. There
are no rules for choosing this distribution, and convenience is the guiding
principle. In this example we choose the prior distribution
in a form that the posterior distribution, \ie, $P(p|{\cal E})$ is of
the same form. This is achieved by the beta distribution,
\beq
   {\rm Pr\/}(p) = \beta(p,a,b) = \frac{\Gamma(a+b)}{\Gamma(a)\,\Gamma(b)}
      p^{a-1}(1-p)^{b-1}.
\eeq{beta}
One has $E[p]=a/(a+b)$ and $V[p]=a(a+1)/(a+b)/(a+b+1)$. The value of
$E[p]$ shows that one should make the identification $a=H_0$ and
$b=N_0-H_0$. After the experiment, the posterior distribution of $p$ is
\beq
\begin{aligned}
   P(p|{\cal E})= \frac{\Gamma(a+b+N)}{\Gamma(a+H)\,\Gamma(b+N-H)}\times \\
      p^{a+H-1}(1-p)^{b+N-H-1}.
\end{aligned}
\eeq{posterior}
The expectation value of $p$ in this distribution is exactly the Bayesian
result given in \eqn{prior}. As a result, maximizing $\log P(p|{\cal E})$
gives the Bayesian result. The interpretation of this is exactly what
we discussed without writing the probability formulae.

An example is shown in \fgn{experiment} with $H=3$ and $N=12$, which
means the frequentist result is $p_1=1/4$. Different Bayesian priors with
$p_0=1/2$ are shown for different $N_0$. With increasing $N_0$ the prior
distribution of $p$ is narrower. The Bayesian posterior distributions
are also shown. When $N_0$ is large, the experimental evidence does not
shift the expectation value of $p$, although the distribution changes
significantly. As $N_0$ becomes small the distributions tend to a limit
shown with a thick black line. This limit is equivalent to the frequentist
result. For $N_0=1$ the expectation value of $p$ is well within the
width of the frequentist distribution of $p$. This is compatible with the
estimate that the error in $p_1$ is of order $1/\sqrt N\simeq0.3$, whereas
the bias due to the Bayesian prior is multiplied by $N_0/N\simeq0.1$.

A generalizable lesson is that choosing an inappropriate prior, $p_0$
need not be a problem. As long as too much reliance is not placed on the
prior (\ie, $N_0$ is chosen small enough), the data can lead to the
empirically supported probability of heads. In other words, bias is
not a problem if overconfidence is avoided.

\subsection{Application to fitting}

The simplest problems of fitting parameters arise with the so-called linear
models, where one has a set of data $C_i$ (with $1\le i\le N$) which
are to be fitted to parameters $p_\alpha$ (with $1\le\alpha\le P$). The
model to be used is $C_i=K_{i\alpha}p_\alpha$ where the coefficient matrix
elements $K_{i\alpha}$ are known. We will collect $C_i$ into a vector $C$
and $p_\alpha$ into a vector $p$. The simplest example is fitting
a straight line to $N$ measurements of data. In this case $P=2$, with
$p_1$ being possibly the intercept and $p_2$ the slope. Then $K_{i1}=1$
for all $i$ and $K_{12}=t_i$ where $t_1$ is the value of the independent
variable at which $C_i$ is measured.

If the covariance matrix of the measurements is $\Sigma$, then the usual
frequentist procedure \cite{lyons} is to maximize the probability that the
parameters describe the data,
\beq
\begin{aligned}
   P(C|p) \propto {\rm e}^{-\chi^2/2},
    \qquad{\rm where}\qquad \\
        \chi^2 = \Delta^T\Sigma^{-1}\Delta, \qquad
        \Delta = C-Kp.
\end{aligned}
\eeq{poster}
Maximizing the posterior probability is the same as minimizing $\chi^2$.
A Bayesian extension is to maximize the prior probability
\beq
   P(p|C) \propto P(C|p) {\rm Pr\/}(p).
\eeq{bayes}
Note the similarity to \eqn{eeq}.

In the general case, two models for the prior probability distribution are
widely used. One corresponding to the Maximum Entropy Method (MEM) is
\beq
   {\rm Pr\/}(p) \propto \exp\left(\gamma\sum_{\alpha=1}^P\left[p_\alpha-p_\alpha^0
    -p_\alpha\log\left(\frac{p_\alpha}{p_\alpha^0}\right)\right]\right).
\eeq{mem}
The parameter $\gamma$ is the single meta-parameter. The larger it is, the 
narrower is the prior distribution. So large values of $\gamma$ correspond
to overconfidence. The other model is used in the Method of Constrained
Fitting (MCF),
\beq
   {\rm Pr\/}(p) \propto \prod_{\alpha=1}^P \exp\left(-\frac{(p_\alpha-p_\alpha^0)^2}{
    2(\sigma_\alpha^0)^2}\right).
\eeq{mcf}
Here there are multiple meta-parameters $\sigma_\alpha^0$. The
smaller they are, the narrower is the prior distribution. So small
values of $\sigma_\alpha^0$ correspond to overconfidence. In view
of the discussion in the preceding subsection, we will dial down the
overconfidence by taking the limit of the results when $\gamma\to0$
or $\sigma_\alpha^0\to\infty$.  We will see later that stability against
changes in the values of these meta-parameters sets in quickly, so that
the limit is not hard to take. Since the meta-parameters appear only in
the prior probability in \eqn{bayes}, exactly the same process can be
used when the parameters appear non-linearly in the fitting function.

\subsection{Application to lattice correlators}

It seems reasonable to argue that Bayesian methods work when the posterior
is not sensitively dependent on priors, as we saw in the toy model before.
This has an effect on the kind of problems which lattice gauge theory can
be used for.

The analytic continuation of thermal (Euclidean) correlators to (Minkowski)
real-time involves an integral relation formally written as
\beq
   C(t) = \int_0^{\infty} d\omega K(\omega,t) \rho(\omega),
\eeq{spectral}
where $C(t)$ are the measured values of the correlator at $0\le t<N_t/2$,
and $K(\omega,t)$ is a known kernel \cite{lgtmem}. Clearly, the extraction
of the function $\rho(\omega)$ is a pathologically under-constrained
problem if one has no prior knowledge of its form.

Assuming zero knowledge, one may discretize the integral into a
Riemann sum over a set $\{\omega_i|1\le i\le M\}$, and write $K_i(t) =
K(\omega_i,t)$ and $\rho_i=(\omega_i-\omega_{i-1})\rho(\omega_i)$ choosing
$\omega_0=0$. Then this becomes a problem of fitting a hyperplane through
a set of data:
\beq
   C(t) = \sum_{i=1}^M \rho_i K_i(t).
\eeq{discrete}
The ordered set $(K_1(t),K_2(t)\cdots,K_M(t),C(t))$ can be thought of as
the coordinate directions in this $M+1$-dimensional space and $\rho_i$
are the slopes of the hyperplane which we want to fit. There are only
$N_t/2+1$ independent pieces of data. When $M>N_t/2+1$, none of the
$\rho_i$ are constrained. Taking a Bayesian approach does not help,
since there is no limit in which the problem is data-driven. As a result,
the prior choices of parameters bias the solution no matter how
the meta-parameters are tuned.

To fix our ideas, we note that taking $N_t=0$ and $M=2$ gives the
problem of fitting a straight line to one piece of data. We know that
any solution to this problem depends on the priors, no matter how the
overconfidence meta-parameters are tuned. Increasing both $N_t$ and
$M$ while keeping $M>N_t/2+1$ does nothing to improve the situation.
The analytic continuation of finite temperature correlators to real-time
has been treated in this formulation.

A careful analysis of the physics may constrain the spectral function
$\rho(\omega)$ in such a way that the problem becomes tractable.  The
results contain priors, but these are vetted by physical constraints. An
example is the argument using a transfer matrix, which says that the
spectral function is a series of Dirac-delta functions. The positions
and strengths of the functions are parameters to be determined.

At zero temperature this form is used along with heuristics which
allow us to use $M<N_t/2+1$. These arise from examining the nature of
$K(\omega,t)$, which predicts an exponential fall in the correlator when
$N_t\to\infty$, and gives rise to the form in \eqn{fitfn}. When $N_t$ is
large enough, then there is a perfectly reasonable likelihood function,
which is able to constrain some of the fit parameters. A Bayesian prior
probability then regulates the problem through the mechanism which we
explored in the analysis of the Bernoulli problem. We explore a borderline
case in this paper where the heuristics begin to break down.

\section{Extracting masses}\label{sec:masses}

\bef
\begin{center}
\includegraphics[scale=0.65]{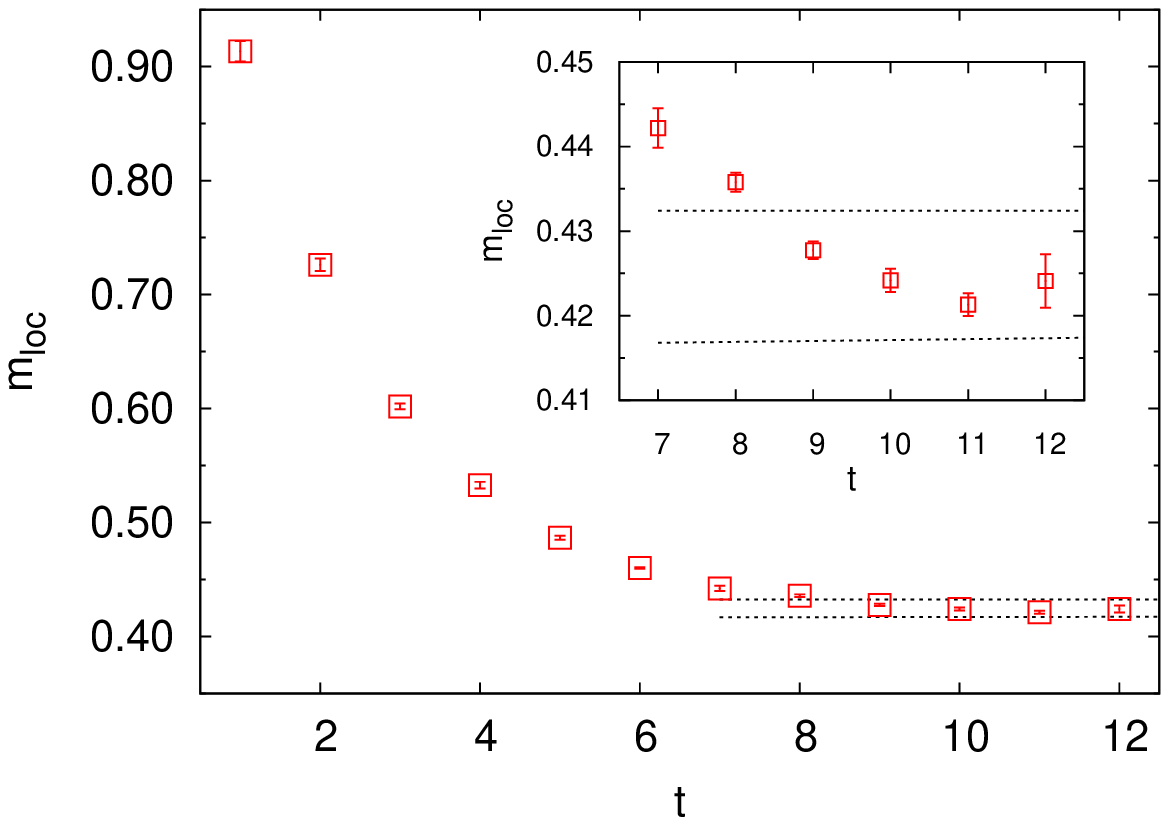}
\end{center}
\caption{Local mass, $m_{loc}$ obtained from a measurement of the pion
correlation function with staggered quarks on $24^4$ lattice with periodic
boundary conditions at $\beta=5.6$ and bare quark mass $ma=0.025$. A
close look shows that there is no plateau in $m_{loc}$ as $t$ changes.}
\eef{localm}

In \cite{ilgti} we had reported a measurement of the pion mass using two
flavours of staggered quarks in a $24^4$ lattice at $\beta=5.6$ and with
bare quark mass $ma=0.025$. The best estimate was $m_\pi a=0.425\pm0.008$.
This was close to, but not in good agreement with, an earlier estimate
\cite{bitar} of $m_\pi a=0.415\pm0.002$ which was made on smaller or
comparable lattices.

However, a closer look at the plot of local masses, shown in \fgn{localm},
shows that the estimate is not very satisfactory, since the local mass
never seems to reach a plateau on lattices of this size. This is odd,
since $m_\pi L \simeq10$, and it would seem that this lattice size should
be more than adequate for the extraction of this mass. This result can
mean that at least one of the excited states cannot be easily decoupled
from the ground state, either because it lies close in mass or because
the operator used to excite a pion couples more strongly to one or more
of these excited state.

In either case it would be possible to create several different kinds of
sources with the same quantum number in order to perform a variational
computation which isolates the ground state. This is the preferred method
today \cite{variation}. However, it is also possible to adapt the MCF
in \cite{bayes} to this problem. In principle this yields a black-box
similar to machine-learning applications today. It would be interesting
also to combine the two methods in future.

The statistical analyses of the Goldstone pion correlators and local
masses are discussed in detail in \cite{ilgti}.  Errors on the local
masses are calculated through a statistical bootstrap, nesting bootstrap
loops where necessary. When the local mass is not constant, then the
approximation of keeping only one state in \eqn{lmass} fails, and one must
keep at least one more mass in the hierarchy of \eqn{fitfn}.  Fitting the
correlator keeping at least two states involves fitting 4 constants.

\bef
\begin{center}
\includegraphics[scale=0.65]{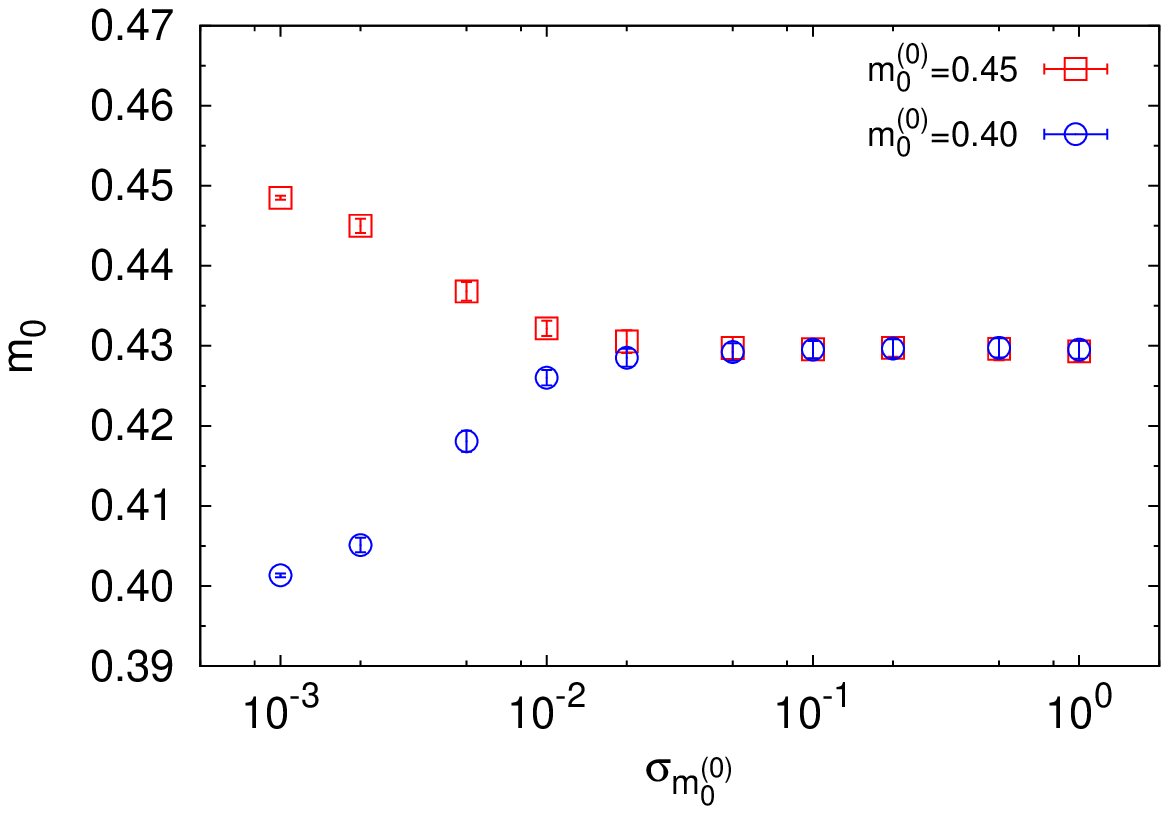}
\includegraphics[scale=0.65]{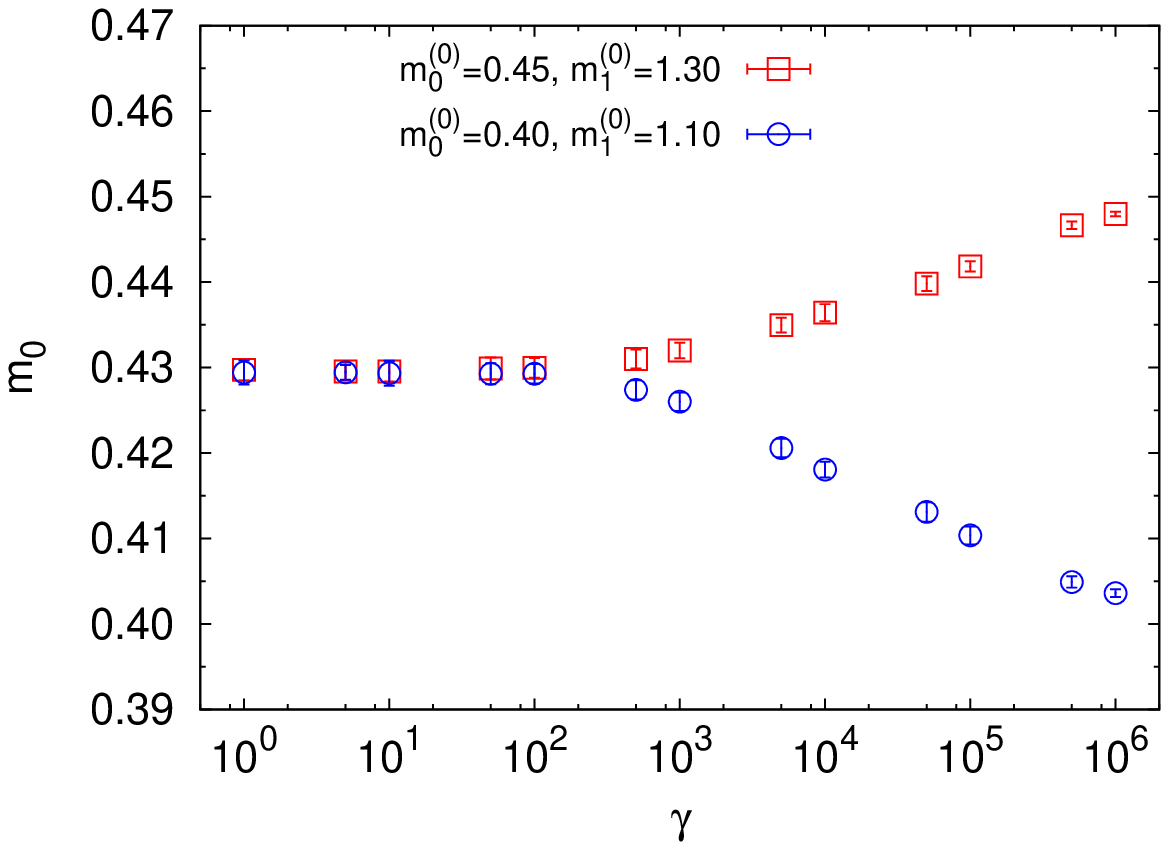}
\end{center}
\caption{(Upper panel) The best fit value of $m_0$ changes systematically
 with the nuisance parameter $\sigma^{(0)}_{m_0}$ in MCF. For small values
 of this parameter the best fit to the mass is determined largely by the
 prior.  However, for larger values, the best fit does not depend on
 $m_0^{(0)}$. (Lower panel) The dependence on the nuisance parameter
 $\gamma$ in MEM is the inverse. Both these figures are for the analysis
 of pion masses at $\beta=5.6$ and bare quark mass $ma=0.025$.}
\eef{stable}

In the MCF we write the prior values $m_0^0$, $m_1^0$, $A_0^0$,
$A_1^0$, and the corresponding meta-parameters $\sigma^0_{m_0}$,
$\sigma^0_{m_0}$, $\sigma^0_{A_0}$, $\sigma^0_{A_1}$ for the parameters
in \eqn{fitfn}. In the MEM, priors carry the same notations as in MCF and
there is a single meta-parameter $\gamma$. We show in \fgn{stable}
that as the meta-parameter is tuned away from overconfidence a stable
fit is obtained that is not very sensitive to the initial bias in the
physical parameters.  We quantify this further in the following way. For
the MCF we choose priors in the large 8-dimensional hypercube with $m_0$
and $m_1$ allowed to vary between 0.1 and 1, $A_0$ varying in the range
from $10^{-3}$ to 0.1, $A_1$ taking values between $10^{-7}$ and $10^{-5}$,
and all the meta-parameters allowed to vary between 0.1 and 10. Within
this hypercube we sampled points using the low-discrepancy Halton sequence
\cite{halton}. We found that a little over 50\% of the volume of priors gave
results with 68\% confidence limits of our best estimate for $m_0$,
more than 95\% of priors yielded a fit value within 95\% confidence limits,
and the full volume yielded results within 99\% confidence limits. Similar
results were obtained in the MEM. This indicates that prior bias is not
a major issue.

Using this method, we have the estimate of the ground state pion mass is
\beq
\begin{aligned}
  & am_0= 0.429\pm0.002 & \\
  & (\beta=5.6,\ am=0.025,\ N_f=2,\ {\rm pion}) &
\end{aligned}
\eeq{pionmass}
with the 68\% confidence limit on it, as shown in \fgn{stable}. Exactly
the same result is obtained on changing the Bayesian analysis from MEM to
MCF. The distribution of the fitted parameter is strongly non-Gaussian,
as can be seen from the fact that the 95\% confidence limits are
$0.429^{+0.052}_{-0.013}$.
The above-calculated value of the ground state mass is consistent with the
previously reported value \cite{bitar}, at the 95\% confidence limit.

We are also able to obtain a similarly stable value for one
excited state
\beq
\begin{aligned}
  & am_1= 1.21\pm0.03 & \\
  & (\beta=5.6,\ am=0.025,\ N_f=2,\ {\rm pion}) &
\end{aligned}
\eeq{expionmass}
at the 68\% confidence level. Note that the difference $a \Delta m =
am_1 - am_0\simeq 0.78$ is large. This implies that the problem in
disentangling the ground state probably comes from the fact that the
composite operator used to excite a pion has a small overlap with the
ground state.

According to the criteria discussed already after \eqn{fitfn}, since
$am_1>1$, we should stop at using only one excited state. The lattice
lacks sensitivity to multiple excited states with masses above the
lattice cutoff. In fact, if we use a third state, with mass $m_2$, and a
corresponding coefficient $A_2$, in \eqn{fitfn}, further complications
arise. The fits tend to $m_1\simeq m_2$ with $A_1$ and $A_2$ varying
wildly. We find that a stable fit can only be obtained if one imposes the
restrictions $m_2\gg m_1$ and $A_2\ll A_1$. This is again an indication
that there are only two masses smaller than or around the inverse lattice
spacing. Scanning the meta-parameters, and priors with the restrictions
above, yields the same stable value of $m_0$.

One way of quantifying the need to include an excited state is to
examine how large a fraction of the correlator at $t=0$ is contained in
the ground state term in \eqn{fitfn}.  Using the measured value, $C(0)$,
of the correlator at $t=0$ and the fitted ground state parameters $am_0$
and $A_0$, we define the ground state overlap as the ratio
\beq
   \Omega = \frac{A_0\cosh(am_0N_t/2)}{C(0)}
\eeq{omega}
In the unlikely case in which $\Omega=1$, the full correlation function,
even at distance $t=0$ would be described by a single state. We find that
$\Omega=0.318\pm0.009$ for this set. Although more than two-thirds of
the correlation function come from excited states, the first excited
state mass is already at the UV cutoff $1/a$, and one cannot resolve
the tower of states above it.

\bef
\begin{center}
\includegraphics[scale=0.65]{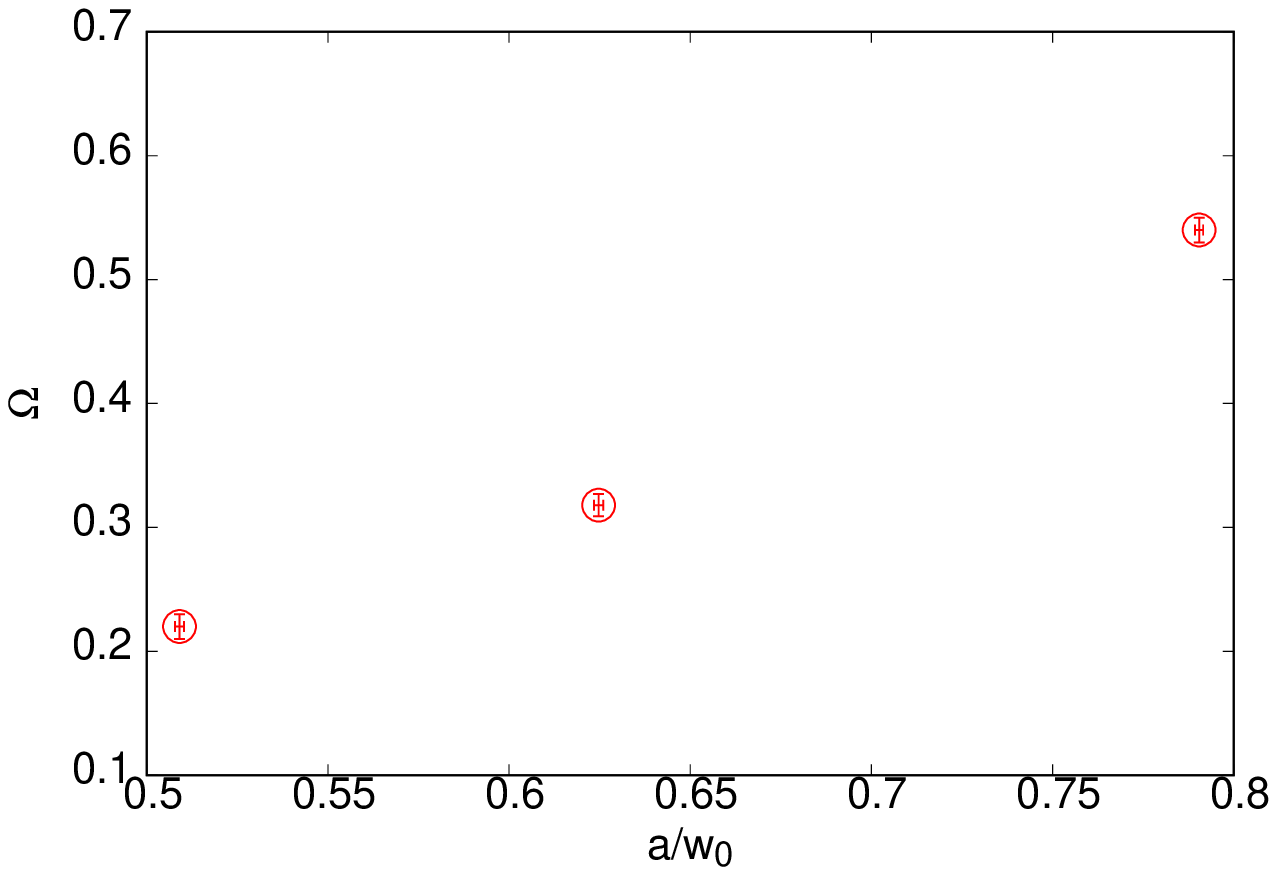}
\end{center}
\caption{The ground state overlap obtained as explained here at $\beta=
5.5$, 5.6 and 5.7 with bare quark mass $ma=0.025$. The lattice spacing
in units of $w_0$ is taken from \cite{ilgti}. With decreasing lattice
spacing the point source has rapidly decreasing overlap with the ground
state.}
\eef{overlap}

We are able to do a similar analysis at $\beta=5.7$ and bare quark mass
$ma=0.025$. Using two terms of the tower in \eqn{fitfn} we find
\beq
\begin{aligned}
  & am_0= 0.407^{+0.005}_{-0.006} & \\
  & (\beta=5.7,\ am=0.025,\ N_f=2,\ {\rm pion}) &
\end{aligned}
\eeq{pionmass2}
where the errors are at the 68\% confidence limit. This data set
gives a more nearly Gaussian distribution of the fitted mass, with
the errors doubling at the 95\% confidence limit.
A previous measurement \cite{brown} gave a similar value namely
$am_\pi=0.388\pm0.001$.

We show the ground state overlap as a function of the lattice spacing
in \fgn{overlap}\footnote{The correlators at $\beta=5.5$ yielded a
clear plateau in local mass. Analysis by the method of this paper gave a
completely compatible result, and the overlap shown in the figure.}. One
sees that a roughly 30\% change in the lattice spacing causes the
overlap to decrease by about a factor of two.  It is interesting that
the same Bayesian analysis can remove the contamination of masses by
higher-lying states even though the overlap decreases so strongly.
Extraction of the pion decay constant using the ground state
amplitude $A_0$ shows that lattice spacing effects are small at fixed
pion mass, when both are extracted in physical units. As a result,
we found that the technique is a good black box even when the
local mass plateau is not fully developed. In these cases we found
that the excited state is non-physical, so the use of the modern
machinery of variational computations \cite{variation} to
simultaneously fit ground and multiple excited states is too ponderous.

\section{Conclusions}\label{sec:conclude}

In this paper we have examined the Bayesian approach to parameter
extraction when all the parameters are not determined by the data. We have
shown that meta-parameters, which we have called overconfidence, cause
the solution to cross over from a prior dominated to a data dominated
region in the best of the cases. When this happens, then a reasonable way
to deal with the meta-parameters is to take them to lie in the region
where the solution is data driven. 

We applied this idea to extracting the lowest mass on
a lattice which is too short for the local masses to show a plateau. This
is usually due to (at least) one other state which has not decoupled since
the lattice is not long enough. Rough estimates of the ground state and
an excited state can often be made from inspection of the data. We showed
that in the multi-parameter space of priors, convergence to a stable
value is obtained once the meta-parameters are fixed using the notions
developed in the previous section. This is, of course, the statement that
the prior bias is not important once the overconfidence parameters have
been dialled down. As a result, there is a black-box method for the fit.


We showed that fitting correlators with \eqn{fitfn} using MCF and MEM
gave results consistent with other published estimates at the 95\%
confidence limits. We have checked that a simple-minded fit to
local masses also gives results consistent with the above-mentioned
estimates. However, the fit to correlators using \eqn{fitfn}
is to be preferred since it makes weaker assumptions about the data.
Our results mildly correct previous measurements of the pion mass
at the same values of bare parameters.

An interesting physics result is given in \fgn{overlap} where we show
quantitatively how the overlap of a point source on the physical pion
decreases with the lattice spacing. Interestingly, we showed that although
the ground state overlap decreases, the excited states which couple
to the operators are above the lattice cutoff. The variational methods
which are used today \cite{variation} to project on to the ground state
are too expensive, since the simultaneous extraction of excited state
properties will not give physics results. In such cases the simple
black-box method that we describe becomes useful.

\begin{acknowledgements}
In the final stage of this work, A.L. is funded from the grant 05P15PBCAA
of the German Bundesministerium f\"ur Bildung und Forschung.
This work used lattice configurations and propagators obtained using the
computing resources of the Indian Lattice Gauge Theory Initiative (ILGTI).
We thank Nikhil Karthik and Pushan Majumdar for their comments.
\end{acknowledgements}


\begin{thebibliography}{0}

\bibitem{bayes}
  D.\ Makovoz, {\sl Nucl.\ Phys.\ Proc.\ Suppl.\/} 53 (1997) 246;\\
  G.\ P.\ Lepage \etal, {\sl Nucl.\ Phys.\ Proc.\ Suppl.\/} 106 (2002) 12;\\
  C.\ Morningstar, {\sl Nucl.\ Phys.\ Proc.\ Suppl.\/} 109A (2002) 185.
\bibitem{jaynes}
   E.\ T.\ Jaynes, {\sl Probability Theory, The Logic of Science\/},
   ed.\ G.\ L.\ Bretthorst, Cambridge University Press, 2015.
\bibitem{bishop}
   C.\ M.\ Bishop, {\sl Pattern Recognition and Machine Learning\/},
   Springer Science+Business Media, Singapore, 2006.
\bibitem{other}
   J.\ R.\ Gubernatis \etal, \pr B 44 (1991) 6011.
\bibitem{feller}
  W.\ Feller, {\sl An Introduction to Probability Theory and its
  Applications\/}, John Wiley and Sons, 3rd Revised Edition (1968).
\bibitem{lyons}
   L.\ Lyons, {\sl Statistics for Nuclear and Particle Physicists\/},
   Cambridge University Press, 1986.
\bibitem{lgtmem}
   Y.~Nakahara, M.~Asakawa and T.~Hatsuda,
  {\sl Phys.\ Rev.\/} D 60 (1999) 091503 [hep-lat/9905034];\\
   T.~Yamazaki {\it et al.} [CP-PACS Collaboration], {\sl Phys.\ Rev.\/}
  D 65 (2002) 014501 [hep-lat/0105030];\\
   S.~Datta, F.~Karsch, P.~Petreczky and I.~Wetzorke,
  {\sl Phys.\ Rev.\/} D 69 (2004) 094507 [hep-lat/0312037].
\bibitem{ilgti}
  S.~Datta, S.~Gupta, A.~Lahiri and P.~Majumdar,
   {\sl Phys.\ Rev.\/} D {\bf 94} (2016) no.5,  054506
  [arXiv:1606.05546 [hep-lat]].
\bibitem{bitar}
  K.~M.~Bitar {\it et al.},
  {\sl Phys.\ Rev.\/} D 42 (1990) 3794.
\bibitem{variation}
  T.~Burch {\it et al.} [Bern-Graz-Regensburg Collaboration],
  Phys.\ Rev.\ D {\bf 70} (2004) 054502
  [hep-lat/0405006];\\
  J.~J.~Dudek, R.~G.~Edwards, N.~Mathur and D.~G.~Richards,
  Phys.\ Rev.\ D {\bf 77} (2008) 034501
  [arXiv:0707.4162 [hep-lat]];\\
  S.~Prelovsek, T.~Draper, C.~B.~Lang, M.~Limmer, K.~F.~Liu, N.~Mathur and D.~Mohler,
  Phys.\ Rev.\ D {\bf 82} (2010) 094507
  [arXiv:1005.0948 [hep-lat]].
\bibitem{halton}
  J.~H.~Halton,
  {\sl Comm.\ A.\ C.\ M.\/}, {\bf 7} (1964) 701.
\bibitem{brown}
  F.\ R.\ Brown, F.\ P.\ Butler, H.\ Chen, N.\ H.\ Christ, Z.\ Dong,
  W.\ Schaffer, L.\ I.\ Unger, and A.\ Vaccarino,
  {\sl Phys.\ Rev.\ Lett.\/} {\bf 67} (1991) 1062.

\end{thebibliography}
\end{document}